\documentclass[a4paper,aps,showpacs,amsmath,amssymb]{revtex4}
\usepackage{mathrsfs}
\usepackage{amsmath}
\usepackage[dvips]{graphicx}
\usepackage{epsfig}
\usepackage[dvips]{graphics,graphicx}
\begin{document}

\title{Opto-mechanical effects in self-organization of a Bose-Einstein condensate in an optical cavity}

\author{ Priyanka Verma$^{1}$ , Aranya B Bhattacherjee$^{2}$ and ManMohan$^{1}$}

\address{$^{1}$Department of Physics and Astrophysics, University of Delhi, Delhi-110007, India} \address{$^{2}$Department of Physics, ARSD College, University of Delhi (South Campus), New Delhi-110021, India}

\begin{abstract}
The influence of mirror motion on the spatial self organization of a Bose-Einstein condensate (BEC) in an optical cavity is studied. We show that the mirror dynamics tends to destroy the process of self organization. An additional external phonon pump is shown to modify the critical photon pump needed to observe the onset of self organization.
\end{abstract}


\maketitle

\section{Introduction}

Bose-Einstein condensates (BEC) are regarded as an ideal coherent system to study many body quantum physics in a highly controlled manner e.g. superfluid-Mott insulator phase transition \citep{greiner}. However this particular kind of phase transition is governed by short range interactions. An interesting example of long range interaction induced phase transition is the creation of a self-organized phase from a BEC in a high $Q$ optical cavity, above a certain critical transverse optical pump intensity \citep{nagy}. At the phase transistion, the spatial symmetry of the cavity optical lattice is spontaneously broken. Infact, this self-organization is equivalent to the Dicke quantum phase transition \cite{emary,baumann} . Numerous theoretical papers have explored novel phases and dynamics of the Dicke model in BEC \citep{bhaseen,liu,nagy2,larson}.

Recently the field of cavity optomechanics has become an attractive research topic with a wide variety of systems ranging from gravitational wave detectors \citep{corbitt1, corbitt2}, nanomechanical cantilevers \citep{hohberger, gigan, arcizet, kleckner, favero, regal}, vibrating microtoroids\citep{carmon, schliesser}, membranes\citep{thompson} and Bose-Einstein condensate \citep{brennecke, murch, bhattacherjee09, bhattacherjee10, treulein, szirmai,hunger2,chen, chiara, steinke, hunger3, chen2, zhang} and atomic ensembles \citep{singh}. A cavity opto-mechanical system, generally consists of an optical cavity with one movable end mirror. Such a system is utilized to cool a micromechanical resonator to its ground state by the pressure exerted by the cavity light field on the movable mirror.
The studies on cavity opto-mechanics of atoms show that sufficiently strong and coherent coupling would enable studies of atom-oscillator entanglement, quantum state transfer, and quantum control of mechanical force sensors. Due to coupling between the condensate wavefunction and the cantilever, mediated by the cavity photons, the cantilever displacement is expected to strongly influence the superfluid properties of the condensate \citep{bhattacherjee09}. Coupled dynamics of a movable mirror and atoms trapped in the standing wave light field of a cavity were studied \citep{meiser}. Over the past few years, experimental investigations made many significant progresses by combining ultracold atoms technology with cavity QED \citep{25,26,27}.
Motivated by these recent developments in the fields of cavity opto-mechanics and self-organization of cold atoms in cavity, this work focuses on the investigation of the opto-mechanical effects in the self-organization of a Bose-Einstein condensate in an optical cavity. In view of the fact that the motion of the cavity mirror significantly changes the intracavity field \citep{bhattacherjee09}, we expect some interesting physics to occur in the self-organization process. We will also focus on the influence of an external force (phonon pump) acting on the movable mirror on the self-organization effect.

\section{The Model}

We consider a cavity quantum optomechanical system consisting of an elongated cigar shaped Bose-Einstein condensate (BEC) at a temperature $T=0$ interacting with  a single mode of a high-Q optical cavity with one movable mirror. The optical cavity is pumped by an external laser of frequency $\omega$ perpendicular to the cavity axis. The pump photons are coherently scattered by the condensate atoms along the cavity axis. Multiple reflections from the cavity mirrors results in the formation of a standing wave which effectively is a one-dimensional optical lattice. For simplicity, we will consider the system dynamics in only one dimension i.e  along the axis of the cavity. Consequently, we will consider only an elongated cigar shaped BEC which can  be created by keeping the frequency of the harmonic trap along the transverse direction much larger than the one along the axial direction (direction of the cavity axis). The atom-pump detuning $\Delta_{A}=\omega-\omega_{A}$ is kept large to suppress the spontaneous emission of photons by the atoms since this is a source of heat, which eventually destroys the condensate. Here $\omega_{A}$ is the atomic transition frequency. We define the cavity-pump detuning as $\Delta_{c}=\omega-\omega_{c}$, where $\omega_{c}$ is the cavity mode frequency. The coupling of the cavity mode with the condensate atoms is characterized by the single-photon Rabi frequency $g$. In addition, the cavity mode is coupled to the movable mirror which is acting like a mechanical oscillator with frequency $\Omega_{m}$. The coupling of the light field to the mirror is via the dimensionless parameter $\epsilon=(x_{0}/\Omega_{m})\frac{d \omega}{d x}|_{x=0}$, $x_{0}=\sqrt{\frac{\hbar}{2 m_{eff} \Omega_{m}}}$ is the zero-point motion of the mechanical mode, $m_{eff}$ is its effective mass. We also assume that there is no direct coupling between the atoms and the mirrors which is possible if the size of the condensate is much smaller than the cavity dimension and the atoms occupy a small region at the center of the cavity.

\begin{figure}[h]
\hspace{-0.0cm}
\includegraphics [scale=0.70]{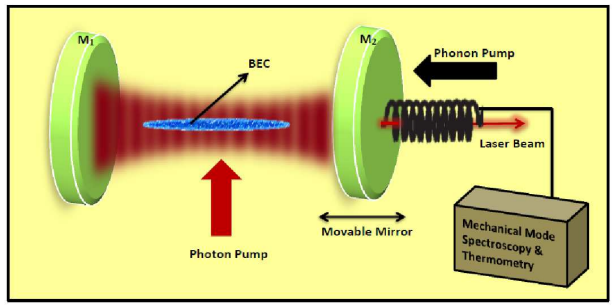}
\caption{Proposed experimental setup of the model to investigate the influence of mirror motion on the self organization of a BEC in an optical cavity.}
\label{f1}
\end{figure}

The condensate is assumed to contain a large number of $N$ atoms with the same wavefunction $\psi(t)$. The cavity field is large and is in a coherent state with a complex amplitude $\alpha$. The phonon field of the mechanical oscillator has a complex amplitude $\alpha_{m}$ with $|\alpha_{m}|^{2}$ as the number of phonons. We will focus on a single mechanical mode by considering a detection bandwidth that includes a single mechanical resonance peak. When the mechanical frequency of the mirror is significantly less than the cavity free spectral range, one can consider only one cavity mode since photon scattering into other modes is neglected. This hybrid system is described by the following coupled mean-field equations:

\begin{equation}
 i \frac{\partial \alpha (t)}{\partial t}=\{ -\Delta_{c}+N<U(x)>+\epsilon \Omega_{m} 2 Re \{\alpha_{m} \}-i\kappa \} \alpha (t)+N <\eta_{t}(x)>
\end{equation}

\begin{equation}
 i \frac{\partial \psi(x,t)}{\partial t}= \{ \frac{p^2}{2\hbar m}+|\alpha(t)|^{2}U(x)+2 Re \{ \alpha(t) \} \eta_{t}(x)+N g_{c}|\psi(x,t)|^2  \} \psi(x,t)
\end{equation}

\begin{equation}
 i \frac{\partial \alpha_{m}(t)}{\partial t}= \{  \Omega_{m}-i \Gamma_{m} \}\alpha_{m}(t)+\epsilon \Omega_{m} |\alpha(t)|^2
\end{equation}

The first equation is the equation for the photon field, the second equation is the Gross-Pitaeviskii equation for the atomic field while
 the last equation is the equation for the phonon field. The condensate and the mirror shifts the cavity resonance frequency by $NU(x)$ and  $\epsilon \Omega_{m} 2 Re \{\alpha_{m}\}$ respectively.
$U(x)=U_{0} \cos^{2}kx$ and $U_{0}=g^2/\Delta_{a}$ is the maximum shift from the cavity resonance due to a single atom. $g$ and $\Delta_{a}$ is the atom-photon coupling strength and the atom-pump detuning respectively.
$<U(x)>=U_{0} <\psi|\cos^{2}kx|\psi>$ is the spatial average over the single atom wavefunction. The transverse pumping is $\eta_{t}(x)= \eta \cos{kx} $ and this term also has to be averaged over the condensate wavefunction. The back action of the light field on the condensate is described by the term $|\alpha(t)|^{2}U(x)$ and on the mirror is described by the term $\epsilon \Omega_{m} |\alpha|^2$ . The system is intrinsically open as the cavity field is damped by the photon leakage and the mechanical oscillator is connected to a bath at finite temperature. Hence $\kappa$ is the optical field damping rate  and $\Gamma_{m}$ is the damping rate of the mechanical oscillator respectively. The two body atom-atom interaction is accounted by the term $g_{c}=4 \pi \hbar a/m w^2$, where $a$ is the s-wave scattering length, $m$ is the mass of a single atom and $w$ is the transverse width of the condensate. In this work, we will work with a large red detuning $U_{0}<0$. This maximizes the light-atom coupling since the atoms are now localized around the intensity maxima.

We now analyze the steady state of the hybrid system. The condensate wavefunction can be written as $\psi(x,t)$ $=$ $\psi_{0}(x) e^{-i \mu t}$, where $\mu$ is the chemical potential. The steady state of the photon, matter wave and the phonon fields are:

\begin{equation}
 \alpha_{0}=\frac{N<\eta_{t}(x)>}{\Delta_{c}-N<U(x)>-\epsilon \Omega_{m} 2 Re \{\alpha_{m,s} \}+i \kappa},
\end{equation}

\begin{equation}
 \{ \frac{p^2}{2 \hbar m}+| \alpha_{0}|^{2}U(x)+2 Re \{ \alpha_{0}\} \eta_{t}(x)+N g_{c}|\psi_{0}(x)|^2  \} \psi_{0}(x)=\mu \psi_{0}(x),
\end{equation}

\begin{equation}
 \alpha_{m,s}=\frac{-\epsilon  \Omega_{m} |\alpha_{0}|^{2}}{\Omega_{m}-i\Gamma_{m}}.
\end{equation}

The steady state values $\alpha_{0}$, $\psi_{0}(x)$ and $\alpha_{m,s}$ has to be determined self consistently using the imaginary time propagation method. Starting from an initial guess, the condensate wavefunction is propagated in imaginary time together with the adiabatic elimination of the cavity and mirror dynamics. The cavity field amplitude and the phonon field amplitude is expressed in terms of the instantaneous wavefunction $\psi(x)$ and inserted into the equation of motion for $\psi(x)$ in each step. The solution that decays slowest is the ground state of the condensate. At each step we have to check the norm $\int dx |\psi(x)|^{2}=1$.   We now re-scale the photon field as $\alpha/\sqrt{N}$ such that the parameters $N U_{0}$, $NG$, $N g_{c}$ and $\sqrt{N} \eta$ is kept constant. All frequency parameters are measured in the unit of recoil frequency $\omega_{R}=\hbar k^{2}/2m$. Substituting the steady state value $\alpha_{m,s}$ in the equation for $\alpha_{0}$, we get a cubic equation in $|\alpha_{0}|^2$:

\begin{equation}
 G^{2} |\alpha_{0}|^6+2 G [\Delta_{c}-N U_{0} B] |\alpha_{0}|^4+[(\Delta_{c}-NU_{0}B)^{2}+\kappa^{2}] |\alpha_{0}^{2}|=N \eta^{2} \varTheta^{2},
\end{equation}

  where, $G=  \dfrac{2 N \epsilon^2 \Omega_{m}^3}{\Omega_{m}^{2}+\Gamma_{m}^{2}}$, the bunching parameter, $B=<\psi_{0}|\cos^{2}kx|\psi_{0}>$, order parameter, $\varTheta=<\psi_{0}|\cos{kx}|\psi_{0}>$. The above equation for the $|\alpha_{0}|$ has to be solved self-consistently since the order parameter and the bunching parameter depends on the value of $|\alpha_{0}|$. Adiabatic elimination of the cavity field leads to the self consistent potential

\begin{equation}
 V(x)= U_{2} \cos^{2}kx+ U_{1}\cos{kx},
\end{equation}

where,

\begin{equation}
U_{1}= 2 \eta Re \{ \alpha_{0} \},
\end{equation}

\begin{equation}
U_{2}= U_{0} |\alpha_{0}|^{2},
\end{equation}

\begin{equation}
 Re \{ \alpha_{0} \}=\frac{N\eta \varTheta [\Delta_{c}-N U_{0} B+G |\alpha_{0}|^2]}{[\Delta_{c}-N U_{0} B+G |\alpha_{0}|^2]^2+\kappa^{2}}.
\end{equation}

\begin{figure}[h]
\hspace{-0.0cm}
\begin{tabular}{cc}
\includegraphics [scale=0.70]{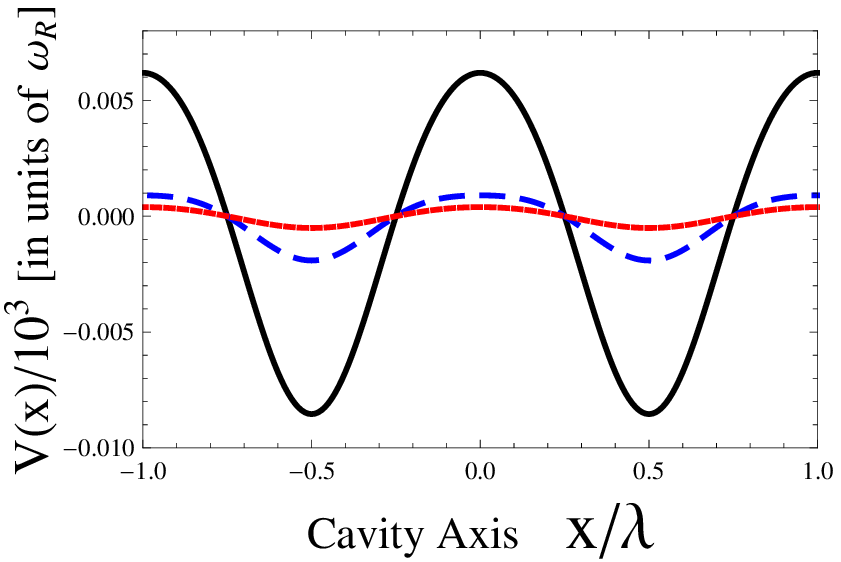}& \includegraphics [scale=0.70] {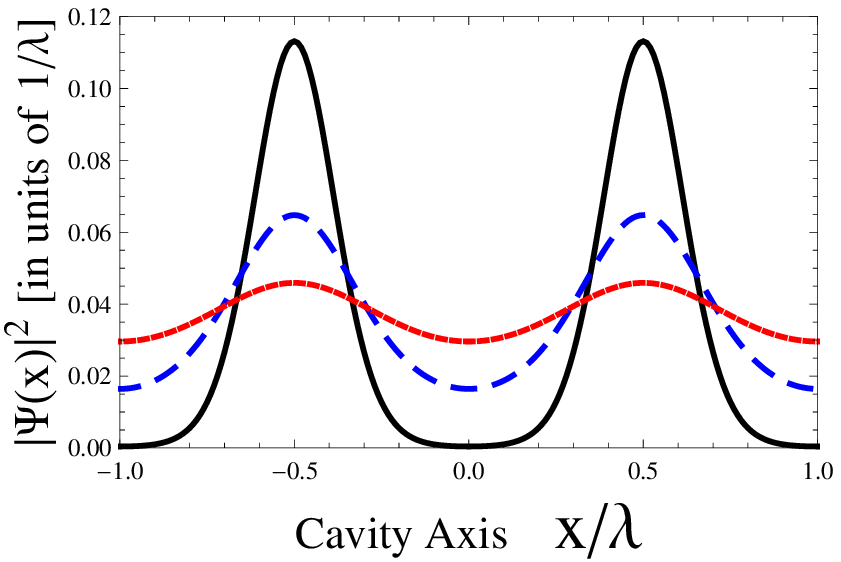}\\
 \end{tabular}
\caption{The effective optical lattice depth (left plot) and the condensate density (right plot) as a function of cavity axis for three values of the photon-mirror coupling, $\epsilon=0$ (solid line), $\epsilon=0.1$ (dashed line) and $\epsilon= 0.3$ (dotted line). The other parameters we consider are: $N=10^4$, $\Omega_{m}/\omega_{R}=40$, $\Gamma_{m}/\omega_{R}=0.0001$, $NU_{0}/\omega_{R}=-100$, $\kappa/\omega_{R}=200$, $\Delta_{c}/\omega_{R}=-300$, $Ng_{c}/\omega_{R}=10 \lambda$ and $\delta_{c}/\omega_{R}=250$. With increasing $\epsilon$, the optical lattice depth decreases and the corresponding density of the BEC at each minima decreases.}
\label{f2}
\end{figure}

\begin{figure}[h]
\hspace{-0.0cm}
\includegraphics [scale=0.70]{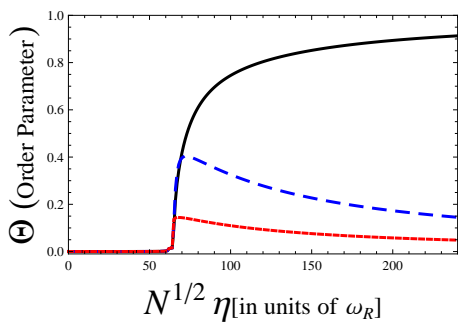}
\caption{Order parameter $\varTheta$, plotted as a function of the transverse pump amplitude $\sqrt{N}\eta$ for three values of photon-mirror coupling, $\epsilon=0$ (solid line), $\epsilon=$ (dashed line) and $\epsilon= $ (dotted line). The other parameters we consider are: $N=10^4$, $\Omega_{m}/\omega_{R}=40$, $\Gamma_{m}/\omega_{R}=0.0001$, $NU_{0}/\omega_{R}=-100$, $\kappa/\omega_{R}=200$, $\Delta_{c}/\omega_{R}=-300$, $Ng_{c}/\omega_{R}=10 \lambda$ and $\delta_{c}/\omega_{R}=250$. An abrupt increase in $\varTheta$ at a critical pump signals the onset of self organization. A finite $\epsilon$ tends to destroy the self organization sice $\varTheta$ is unable to reach the value of unity which implies a complete self organized phase.}
\label{f3}
\end{figure}

 The main influence of the mirror motion is the reduction of the effective optical lattice depth. With progressively increasing mirror-photon coupling, the effective optical lattice depth decreases as evident from the left plot of Fig.2. The resolved side band regime i.e $\Delta_{c}<0$ leads to mirror cooling \citep{naeini}, which leads to a decrease in $\alpha_{m,s}$. A stronger photon-mirror coupling further decreases the mirror amplitude. An inspection of Eqn.(6) reveals that with increasing $\epsilon$, the photon number $|\alpha_{0}|^2$ and hence the optical potential $V(x)$ decreases. As the lattice depth decreases, the condensate density at each minima decreases as shown in the right plot of Fig.2. In Fig.3, we show the influence of mirror motion on the self organization of the BEC in the optical lattice. The appearance of the self organized phase of the BEC in the cavity optical lattice is evident from the variation of the order parameter $\varTheta$ against the pumping strength $\sqrt{N} \eta$. An abrupt increase in $\varTheta$ is an indication of the self organization process. As the mirror-photon coupling increases, the condensate is unable to localize itself at the even sites. Rather the condensate atoms are re-distributed amongst the even and odd sites due to the mirror motion. We could say that the mirror motion spills the atoms out of the even sites.  However, the onset of self organization at a critical pump power does not change with mirror motion.
Let us now understand the influence of the mirror motion on the self organization process.  In the self organization process, there is an abrupt change in the self consistent steady state at a critical laser pump power $\eta_{cr}$. The atoms self-organize as a result of positive feedback from the self-consistent potential part $2 Re \{\alpha_{0} \} \cos {kx}$, which takes a minima either on the even site ($\varTheta >0 $) or the odd site ($\varTheta < 0$).For example, if density fluctuations of the condensate induces $\varTheta > 0$ and if in the absence of the mirror motion, $\Delta_{c} < -N |U_{0}|$, the optical lattice potential resulting from light scattering further pulls the condensate atoms towards the even sites ($kx=2n \pi$) due to a minima at the even sites. This in turn enhances the light scattering from the pump into the cavity and hence a runaway process is initiated. The system reaches a steady state when the gain in potential energy is balanced by the kinetic energy and collision process. Now in the case with mirror motion, the self consistent solution can be found from the real roots of the Eqn.(7) and the matter wave equation (2). Now due to mirror motion the condition necessary for the self-organization process is modified as $\Delta_{c}+ G|\alpha_{0}|^{2} <-N |U_{0}|$. This indicates that the function of the mirror is to counter act the self-organization process. The $\lambda$ periodic potential $U_{1}=2 Re{\alpha_{0}} \cos {kx}$ must hence have a sign opposite to that of $\varTheta$ for the self-organization process to take place. The other $\lambda/2$ periodic $\cos^{2}{kx}$ potential is normally small and its influence is disregarded. The $\cos^{2}{kx}$ has a minima both at the odd as well as the even sites and thus counteracts the self organization process. The ratio $U_{2}/U_{1}$ is given as:

 \begin{equation}
 \frac{U_{2}}{U_{1}}=\frac{\varTheta N U_{0}}{2 [\Delta_{c}-N U_{0} B+G |\alpha_{0}|^{2}]}
 \end{equation}

 The mirror motion term $G |\alpha_{0}|^{2}$ acts in a manner to counter the self organization process. In the presence of the mirror motion, the ratio $U_{2}/U_{1}$ increases making the potential $U_{2}$ more significant and thereby decreasing the self organizing effect. This results in a decrease in the order parameter $\varTheta$. An increase in the pump localizes the atoms in both the odd and even sites, decreasing $\varTheta$ even further and for large $\sqrt{N} \eta_{}$, $\varTheta \rightarrow$ $0$ (uniform atomic distribution). However, the onset of self organization at a critical pump power does not change with mirror motion. This can be understood from the expression for the critical pump calculated analytically for the below threshold case of $\psi_{0}(x)=1$, $\alpha_{0}=0$, $\alpha_{m,s}=0$ and $\mu = Ng_{c}$.

 \begin{equation}
 \sqrt{N} \eta_{c}= \sqrt{\frac{(\Delta_{c}-NU_{0}/2)^{2}+\kappa^{2}}{(NU_{0}-2 \Delta_{c})}} \sqrt{\omega_{R}+2 N g_{c}}
 \end{equation}

Interestingly, below threshold the mirror has no influence on the critical pump value since the mirror motion term does not appear in Eqn (11).

Fig.4 (left plot) shows the steady state value of the cavity photon number as a function of the pump. An abrupt increase in the number of photons is seen at the critical pump value as found experimentally in the paper of Esslinger group \citep{baumann}. Note that self-organization is manifested by an abrupt build-up of the cavity field accompanied by the formation of momentum component of the BEC \citep{baumann}. As the mirror photon coupling increases, the steady state value of the cavity photon decreases. As noted earlier, a finite coupling between the cavity field and the mirror leads to a reduction in the strength of the cavity field. Steady state displacement of the mirror as a function of the pump intensity shown in Fig.4 (right plot). An abrupt increase in the steady state mirror displacement is observed at the critical pump. A decrease in the cavity field due to finite mirror-photon coupling naturally leads to a decrease in the radiation pressure and hence a decrease in the mirror displacement. The continuous monitoring of the mirror motion could serve as an alternate tool to observe the Dicke phase transition (onset of self organization). The displacement of the mirror decreases with increasing value of the photon-mirror coupling.

\begin{figure}[h]
\hspace{-0.0cm}
\begin{tabular}{cc}
\includegraphics [scale=0.70,angle=0]{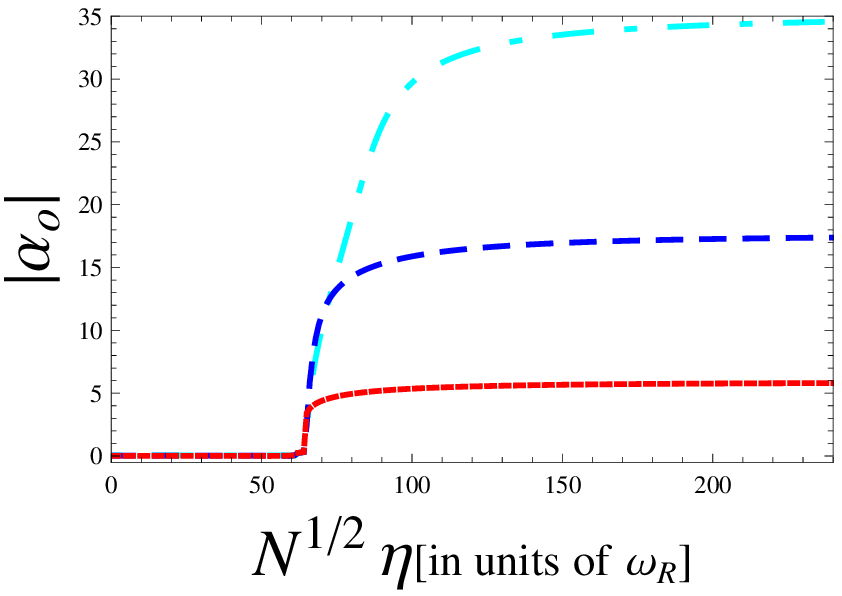}& \includegraphics [scale=0.70] {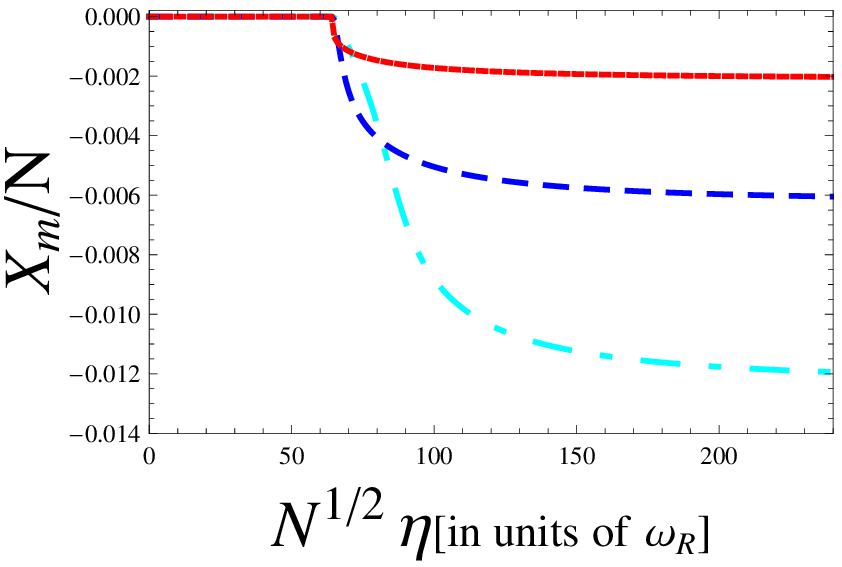}\\
\end{tabular}
\caption{Steady state value of the cavity photon number (left plot) and displacement of the mirror (right plot) as a function of the pump. An abrupt increase is seen at the critical pump value as found experimentally in the paper of Esslinger group. As the mirror photon coupling increases ($\epsilon=0.05$ (dot-dashed line), $\epsilon=0.1$ (dashed line) and $\epsilon= 0.3$ (solid line)), the steady state value of the cavity photon decreases. An abrupt increase in the mirror displacement is observed at the critical pump. This could serve as an alternate mechanism to observe the Dicke phase transition. The displacement of the mirror decreases with increasing value of the photon-mirror coupling. The other parameters we consider are: $N=10^4$, $\Omega_{m}/\omega_{R}=40$, $\Gamma_{m}/\omega_{R}=0.0001$, $NU_{0}/\omega_{R}=-100$, $\kappa/\omega_{R}=200$, $\Delta_{c}/\omega_{R}=-300$, $Ng_{c}/\omega_{R}=10 \lambda$ and $\delta_{c}/\omega_{R}=250$.}
\label{f4}
\end{figure}

\section{Spectrum of small fluctuations}

 We now study the excitation spectrum of the coupled condensate-mirror system as the linear response of the cavity field self-consistent steady stae.The collective excitation spectrum of the hybrid system i.e coupling of the fluctuation of the mechanical oscillator, cavity field and the condensate leads to normal mode splitting \citep{bhattacherjee09}. This optomechanical spectrum is a result of driving three parametrically coupled non-degenerate modes out of equilibrium. To this end, we need to consider the deviations from the steady state i.e $\psi_{0}$, $\alpha_{0}$ and $\alpha_{m,s}$.

\begin{equation}
\alpha(t)=\alpha_{0}+\delta \alpha(t),
\end{equation}

\begin{equation}
\psi(x,t)=e^{-i\mu t}[\psi_{0}(x)+\delta \psi(x,t)],
\end{equation}

\begin{equation}
\alpha_{m}(t)= \alpha_{m,s}+\delta \alpha_{m}(t).
\end{equation}

The resulting linearized equation of motion are,

\begin{eqnarray}
i \frac{d \delta \alpha(t)}{dt} &=& A \delta \alpha(t)+N [<\psi_{0}|\eta_{t}(x)|\delta \psi>+ <\delta \psi|\eta_{t}(x)|\psi_{0}>]+N \alpha_{0}[<\psi_{0}|U(x)|\delta \psi> \\  \nonumber &+& <\delta \psi|U(x)|\psi_{0}>]+2 \epsilon \Omega_{m} [Re (\delta \alpha_{m}(t))\alpha_{0}+Re(\alpha_{m,s}) \delta \alpha (t)],
\end{eqnarray}

\begin{eqnarray}
i \frac{d \delta \psi(x,t)}{dt} &=& {H_{0}+Ng_{c}|\psi_{0}(x)|^{2}} \delta \psi (x,t)+N g_{c} \psi_{0}^{2}(x) \delta \psi^{*}(x,t)\\ \nonumber &+& \psi_{0}(x)U(x)[\alpha_{0} \delta \alpha^{*}(t)+\alpha_{0}^{*} \delta \alpha(t)]+\psi_{0}(x) \eta_{t}(x) [\delta \alpha(t)+\delta \alpha^{*}(t)],
\end{eqnarray}

\begin{equation}
i \frac{d \delta \alpha_{m}(t)}{dt}=(\Omega_{m}-i \Gamma_{m}) \delta \alpha_{m}(t)+ \epsilon \Omega_{m}[\delta \alpha(t) \alpha^{*}_{0}+\delta \alpha^{*}(t) \alpha_{0}],
\end{equation}

where $A=-\Delta_{c}+N <\psi_{0}|U(x)|\psi_{0}>- i \kappa$, and $H_{0}=p^{2}/2m+N g_{c} |\psi_{0}^{2}|^{2}-\mu +|\alpha_{0}|^{2} U(x)+2 Re \{\alpha_{0}\}\eta_{t}(x)$.

The solution of the above coupled equations can be looked in the form,

\begin{equation}
\delta \alpha(t)= e^{-i \omega t} \delta \alpha_{+}+ e^{i \omega^{*}t} \delta \alpha_{-}^{*},
\end{equation}

\begin{equation}
\delta \psi(x,t)= e^{-i \omega t} \delta \psi_{+}(x)+ e^{i \omega^{*}t} \delta \psi_{-}^{*}(x),
\end{equation}

\begin{equation}
\delta \alpha_{m}(t)=e^{-i \omega t} \delta \alpha_{m,+} + e^{i \omega^{*} t} \delta \alpha_{m,-}^{*},
\end{equation}

where, $\omega=\nu-i\gamma$ is the complex frequency with damping rate $\gamma$. It is convenient to solve the coupled equations in the following symmetric and anti-symmetric basis, $\delta \alpha_{a}=\delta \alpha_{+}-\delta \alpha_{-}$, $\delta \alpha_{s}=\delta \alpha_{+}+\delta \alpha_{-}$, $\delta \alpha_{m,a}=\delta \alpha_{m,+}-\delta \alpha_{m,-}$, $\delta \alpha_{m,s}=\delta \alpha_{m,+}+\delta \alpha_{m,-}$, $\delta g_{x}=\delta \psi_{-}(x)-\delta \psi_{+}(x)$, $\delta f_{x}=\delta \psi_{-}(x)+\delta \psi_{+}(x)$. In the above symmetric and anti-symmetric basis, the linear eigenvalue equation is

\begin{equation}
\omega \left(  \begin{array}{c} \delta \alpha_{a}\\ \delta \alpha_{0}\\ \delta f(x)\\ \delta g(x)\\ \delta \alpha_{m,a}\\ \delta \alpha_{m,s} \end{array} \right )= \textbf{M} \left(  \begin{array}{c} \delta \alpha_{a}\\ \delta \alpha_{0}\\ \delta f(x)\\ \delta g(x)\\ \delta \alpha_{m,a}\\ \delta \alpha_{m,s} \end{array} \right ),
\end{equation}

where the non-Hermitian matrix $\textbf{M}$ is,

\begin{equation}
\textbf{M}=\left( \begin{array}{cccccc} -i \kappa & \delta_{cm} & 2N[Re(\alpha_{0})X+Y] & 0 & 0 & 2 \epsilon \Omega_{m} Re (\alpha_{0})\\ \delta_{cm} & -i \kappa & 2 i N X Im (\alpha_{0}) & 0 & 0& 2i\epsilon \Omega_{m} Im(\alpha_{0})\\ 0 & 0 & 0& -H_{0} & 0 & 0\\ 2 i \psi_{0} U Im(\alpha_{0}) & -2 \psi_{0}[U Re(\alpha_{0}+ \eta_{t})] & -H_{0}-2 N g_{c} \psi_{0}^{2} & 0 & 0 & 0\\ -2 i \epsilon \Omega_{m} Im(\alpha_{0}) & 2 \epsilon \Omega_{m} Re(\alpha_{0}) & 0 & 0 & -i \Gamma_{m} & \Omega_{m}\\ 0 & 0 & 0 & 0 & \Omega_{m} & -i \Gamma_{m}   \end{array}    \right)
\end{equation}

Here, $\delta_{cm}= (-\Delta_{c}+\frac{N U_{0}}{2}+2 \epsilon \Omega_{m} Re{\alpha_{m,s}})$ and $X \xi(x)= \int dx  \psi_{0}(x) U(x) \xi(x)$ and $Y \xi(x)= \int dx \psi_{0}(x) \eta_{t}(x)  \xi(x)$ are the overlap integrals. We now calculate the collective excitation spectrum below the critical point of self organization and analyze how the mirror dynamics influences the spectrum. We assume a homogeneous atomic distribution $\psi_{0}=1$ and vanishing cavity field $\alpha_{0}=0$ and phonon field $\alpha_{m,s}=0$ below threshold. This immediately reduces the matrix $\textbf{M}$.

\begin{equation}
\textbf{M}=\left( \begin{array}{cccccc} -i \kappa & \delta_{c} & 2NY & 0 & 0 & 0\\ \delta_{c} & -i \kappa & 0 & 0 & 0& 0 \\ 0 & 0 & 0& -\omega_{R} & 0 & 0\\ 0 & -2 \eta_{t} & -\omega_{R}-2 N g_{c}  & 0 & 0 & 0\\ 0 & 0 & 0 & 0 & -i \Gamma_{m} & \Omega_{m}\\ 0 & 0 & 0 & 0 & \Omega_{m} & -i \Gamma_{m}   \end{array}    \right),
\end{equation}

where, $\delta_{c}= (-\Delta_{c}+\frac{N U_{0}}{2})$.

This has the sixth order characteristic equation,

\begin{equation}
[(i \Gamma_{m}+\lambda)^{2} - \Omega_{m}^{2}] \{ [(i \kappa+\lambda)^{2}-\delta_{c}^{2}](\lambda^{2}-\Omega_{1}^{2})-2 N \eta^{2} \Omega_{R} \delta_{c} \}=0,
\end{equation}

where $\Omega_{1}=\sqrt{\omega_{R}(\Omega_{R}+2 N g_{c})}$ is the first excitation energy in the Bogoliubov spectrum of a BEC. In the absence of the pump ($\eta=0$), the three excitations decouple i.e $\lambda_{1,2}=\pm \Omega_{1}$ for the condensate excitation, $\lambda_{3,4}=\pm \delta_{c}-i \kappa$ for the cavity mode and $\lambda_{5,6}=\pm \Omega_{m}-i \Gamma_{m}$ for the mirror mode. For a finite pumping the cavity mode and the condensate mode couples but the mirror mode still remains decoupled. In this case the mirror motion does not influence the collective excitations below threshold. A full coupling between all the three modes can be achieved only for a finite steady state mirror motion, $\alpha_{m,s}$, which we show in the next section that it can be achieved with a phonon pump.

\section{Dynamics in the presence of phonon pump}

In this section, we consider the dynamics of self organization in the presence of a phonon pump i.e an external source which excites the mirror. This source could be in a form of a mechanical object in physical contact with the mirror or an external laser which vibrates the mirror due to radiation pressure. The external pump will couple with the amplitude quadrature of the mirror fluctuations. The equation of motion for the mirror is rewritten in the presence of the phonon pump as,

\begin{equation}
 i \frac{\partial \alpha_{m}(t)}{\partial t}= \{  \Omega_{m}-i \Gamma_{m} \}\alpha_{m}(t)+\epsilon \Omega_{m} |\alpha(t)|^2+\eta_{m}.
\end{equation}

Here $\eta_{m}$ is the phonon pump frequency. As a result the steady state of the mirror is modified as

\begin{equation}
Re(\alpha_{m,s})=-\frac{(\epsilon \Omega_{m}^{2}|\alpha_{0}|^2+\eta_{m} \Omega_{m})}{\Omega_{m}^{2}+\Gamma_{m}^{2}}.
\end{equation}

\begin{figure}[h]
\hspace{-0.0cm}
\includegraphics [scale=0.55] {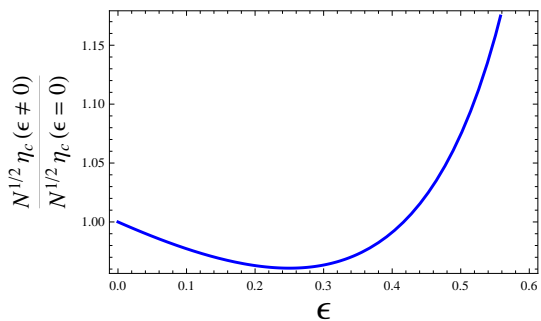}
\caption{The ratio $\frac{(\sqrt{N}\eta_{c})_{\epsilon\neq 0}}{(\sqrt{N}\eta_{c})_{\epsilon=0}}$ is plotted as function of the mirror-cavity field coupling, $\epsilon$. Parameters we consider are: $\eta_{m}/\omega_{R}=100$, $N=10^4$, $\Omega_{m}/\omega_{R}=40$, $\Gamma_{m}/\omega_{R}=0.0001$, $NU_{0}/\omega_{R}=-100$, $\kappa/\omega_{R}=100$, $\Delta_{c}/\omega_{R}=-200$, $Ng_{c}/\omega_{R}=10 \lambda$ and $\delta_{c}/\omega_{R}=150$.}
\label{f5}
\end{figure}

This immediately changes the expression for the critical pump below threshold as,

\begin{equation}
 \sqrt{N} \eta_{c}= \sqrt{\frac{[\Delta_{c}-NU_{0}/2-(2\epsilon \Omega_{m}^{2} \eta_{m})/(\Omega_{m}^{2}+\Gamma_{m}^{2})]^{2}+\kappa^{2}}{(NU_{0}+(4\epsilon \Omega_{m}^{2} \eta_{m})/(\Omega_{m}^{2}+\Gamma_{m}^{2})-2 \Delta_{c})}} \sqrt{\omega_{R}+2 N g_{c}}
 \end{equation}

The ratio $\frac{(\sqrt{N}\eta_{c})_{\epsilon\neq 0}}{(\sqrt{N}\eta_{c})_{\epsilon=0}}$ is plotted in fig 5  as function of the mirror-cavity field coupling, $\epsilon$. The critical pump shows an initial decrease and then increase with $\epsilon$ in the presence of the external phonon pump. For $\epsilon<0.4$, the onset of self organized phase occurs below the $(\sqrt{N}\eta)_{\epsilon=0}$, while for $\epsilon > 0.4$, the onset of self organized phase occurs above the critical pump for $\epsilon = 0$.

\begin{figure}[h]
\hspace{-0.0cm}
\includegraphics [scale=0.75]{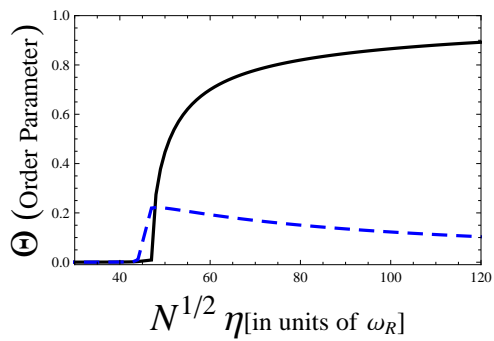}
\caption{Order parameter $\varTheta$, plotted as a function of the transverse pump amplitude $\sqrt{N}\eta$ in the presence of a phonon pump $\eta_{m}$ for two values of photon-mirror coupling, $\epsilon=0.0$ (solid line), $\epsilon= 0.1$ (dashed line).Parameters we consider are: $\eta_{m}/\omega_{R}=100$, $N=10^4$, $\Omega_{m}/\omega_{R}=40$, $\Gamma_{m}/\omega_{R}=0.0001$, $NU_{0}/\omega_{R}=-100$, $\kappa/\omega_{R}=100$, $\Delta_{c}/\omega_{R}=-200$, $Ng_{c}/\omega_{R}=10 \lambda$ and $\delta_{c}/\omega_{R}=150$.}
\label{f6}
\end{figure}

Fig.6 illustrates the order parameter $\varTheta$, as a function of $\sqrt{N}\eta$ for different mirror photon coupling ($\epsilon=0$ (solid line), $\epsilon= 0.1$ (dashed line)) in the presence of the phonon pump. As expected from the analytic expression, the critical pump amplitude ($\sqrt{N} \eta$) changes with $\epsilon$. In the presence of the phonon pump, the drop in the value of $\varTheta$ is much faster as compared to that without the external phonon pump.

The collective excitation spectrum of the coupled condensate-cavity-mirror system also changes in the presence of the external phonon pump. The characteristic equation below threshold is

\begin{equation}
[(i \Gamma_{m}+\lambda)^{2} - \Omega_{m}^{2}] \{ [(i \kappa+\lambda)^{2}-\delta_{cm}^{2}](\lambda^{2}-\Omega_{1}^{2})-2 N \eta^{2} \Omega_{R} \delta_{cm} \}=0,
\end{equation}

where $\delta_{cm}=(-\Delta_{c}+N U_{0}/2+(2 \epsilon \Omega_{m}^{2} \eta_{m})/(\Omega_{m}^{2}+\Gamma_{m}^{2})$. A complete mixing of the three modes i.e the condensate mode, cavity mode and the mirror mode now occurs. A coherent energy exchange between the three modes now takes place, which can be controlled by $\eta_{m}$. A plot of the eigenvalues for different $\epsilon$ for a finite $\eta_{m}$ is shown in Fig.6. Clearly the real part of the eigenvalues goes to zero at the critical pump amplitude and it changes with $\epsilon$ due to a finite $\eta_{m}$.

\begin{figure}[h]
\hspace{-0.0cm}
\begin{tabular}{cc}
\includegraphics [scale=0.70]{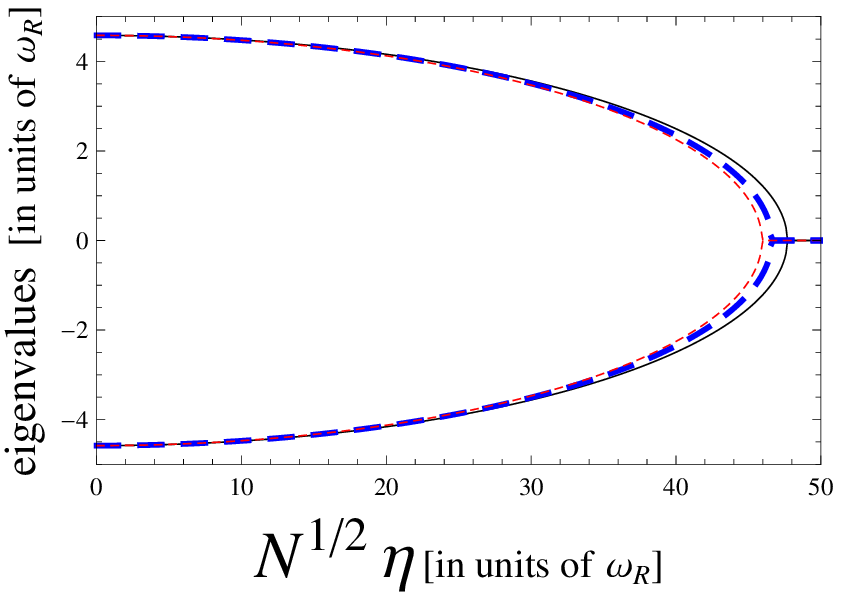}& \includegraphics [scale=0.70] {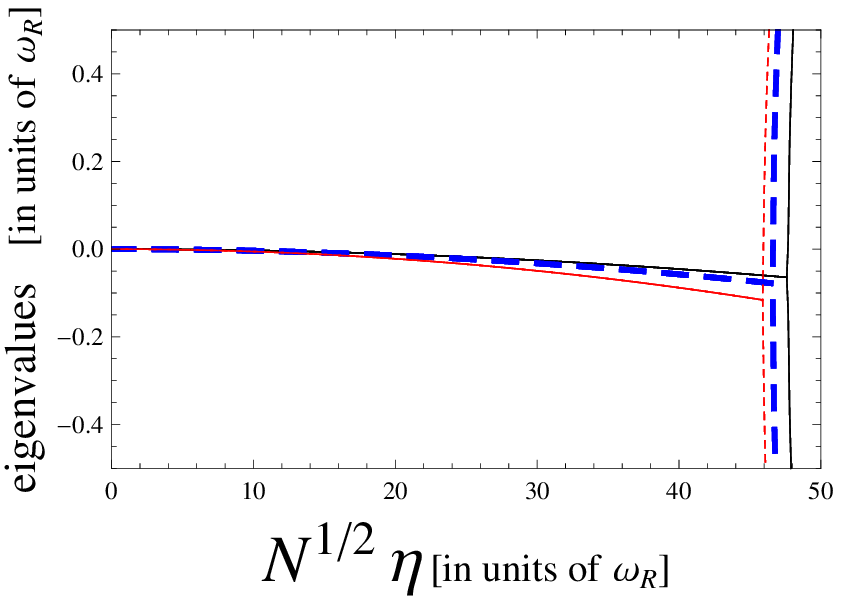}\\
 \end{tabular}
\caption{Real part(left plot) and imaginary part (right plot) of the eigenvalues of the lowest BEC excitation of a homogeneous BEC from Eqn.(28) as a function of the pump strength $\sqrt{N} \eta$ for $\epsilon=0$ (solid line), $\epsilon= 0.1$ (dashed line) and $\epsilon= 0.3$ (dotted line). The real part vanishes at the point where the imaginary parts abruptly increases.Parameters we consider are: $\eta_{m}/\omega_{R}=100$, $N=10^4$, $\Omega_{m}/\omega_{R}=40$, $\Gamma_{m}/\omega_{R}=0.0001$, $NU_{0}/\omega_{R}=-100$, $\kappa/\omega_{R}=100$, $\Delta_{c}/\omega_{R}=-200$, $Ng_{c}/\omega_{R}=10 \lambda$ and $\delta_{c}/\omega_{R}=150$.}
\label{f7}
\end{figure}

The imaginary part of the eigenvalues of Eqn.(28) describes damping which arises from the non-adiabaticity of the cavity field and mirror dynamics i.e the cavity field and the mirror follows the BEC wave-function with a delay of the order of $1/\kappa$ and $1/\Gamma_{m}$ respectively. At the onset of self-organization, the uniform ground state $\psi_{0}$ changes which appears as the vanishing of the real part of the eigenvalues calculated from Eqn.(28), accompanied by a sudden appearance of a positive and negative imaginary part of the eigenvalue. This implies the instability of the condensate \citep{nagy}. The important point here is the shift in the critical pump due to finite $\epsilon$ in the presence of the phonon pump. Usually the movable mirror is used as a ponder-motive detector to measure weak forces acting on it \citep{vitali}. The phonon pump considered here could be a weak force which could be measured from the position of the critical point by appropriately calibrating the device.

\section{Conclusions}
We have studied the effect of a moving cavity mirror on the self-organization, eigenvalue spectrum of a Bose-Einstein condensate confined in an optical cavity. It is found that the mirror dynamics tends to inhibit the self-organization process. We also propose that the continuous monitoring of the mirror displacement could serve as an alternate tool to observe the Dicke super-radiance which is the signature of self-organization. Further we find that the critical laser pump intensity needed to initiate the self-organization process can be coherently controlled by an external phonon pump.This system could serve as a new quantum device to measure weak forces.

\section{Acknowledgements}

One of the authors Priyanka Verma thanks the University Grants Commission, New Delhi for the Junior Research Fellowship. A. Bhattacherjee acknowledges financial support from the Department of Science and Technology, New Delhi for financial assistance vide grant SR/S2/LOP-0034/2010.

\end{document}